\documentclass[conference]{IEEEtran}
\IEEEoverridecommandlockouts
\usepackage{cite}
\usepackage{amsmath,amssymb,amsfonts}
\usepackage{algorithmic}
\usepackage{graphicx}
\usepackage{textcomp}
\usepackage{xcolor}
\usepackage{amsthm}
\usepackage{algorithm}
\usepackage{booktabs}
\usepackage[outdir=./]{epstopdf}
\usepackage{subfigure}
\usepackage{subeqnarray}
\usepackage{bm}
\usepackage{amsfonts}
\usepackage{amssymb}
\usepackage{makecell}
\usepackage{calc}
\usepackage{color}
\usepackage{mathrsfs}
\usepackage{flushend}
\usepackage{url}
\usepackage{stfloats}

\def\BibTeX{{\rm B\kern-.05em{\sc i\kern-.025em b}\kern-.08em
    T\kern-.1667em\lower.7ex\hbox{E}\kern-.125emX}}
\begin{document}

\title{Enhancing the Physical Layer Security of Dual-functional Radar Communication Systems\\
}

\author{\IEEEauthorblockN{ Nanchi Su, Fan Liu, Christos Masouros}
\IEEEauthorblockA{Department of Electronic and Electrical  Engineering,
University College London,
London, UK \\
\{nanchi.su.18, fan.liu, c.masouros\}@ucl.ac.uk,}
}

\maketitle

\begin{abstract}
Dual-functional radar communication (DFRC) system has recently attracted significant academic attentions as an enabling solution for realizing radar-communication spectrum sharing. During the DFRC transmission, however, the critical information could be leaked to the targets, which might be potential eavesdroppers. Therefore, the physical layer security has to be taken into consideration. In this paper, fractional programming (FP) problems are formulated to minimize the signal-to-interference-plus-noise ratio (SINR) at targets under the constraints for the SINR of legitimate users. By doing so, the secrecy rate of communication can be guaranteed. We first assume that communication CSI and the angle of the target are precisely known. After that, problem is extended to the cases with uncertainty in the target's location, which indicates that the target might appear in a certain angular interval. Finally, numerical results have been provided to validate the effectiveness of the proposed method showing that it is viable to guarantee both radar and secrecy communication performances by using the techniques we propose.
\end{abstract}


\section{Introduction}
With the rapid growth of the wireless communication industry, the requirement for extra spectrum resources is on the rise, which motivates the network providers and the policy regulators to seek the chance for reusing frequency bands that are currently under-utilized by the radar systems \cite{rihan2018optimum}. To this end, the research of radar-communication coexistence (RCC) has recently attracted considerable attention from both academia and industry \cite{7898445, mahal2017spectral, liu2019joint}. While various techniques have been proposed for realizing RCC, one crucial drawback is that most of these approaches require frequent cooperation between radar and communication systems, which might be difficult to be implemented in practice. Therefore, a more promising methodology would be to share both the spectrum and the hardware platform between these two functionalities, which requires no additional coordinations, and has motivated the study of the dual-functional radar-communication (DFRC) system \cite{8386661, 8355705, 8737000}. In \cite{hassanien2016dual}, the authors proposed to embed communication symbols into the sidelobes of the radar transmit beampattern, with the mainlobe being employed for target detection, which allows information delivery to single or multiple communication directions outside the mainlobe of the radar. To enable the DFRC transmission for non-line-of-sight (NLoS) communication channels, a joint beamforming design has been presented by \cite{8288677} for simultaneous target detection and multi-user communications, which aims at approaching a desired radar beampattern while guaranteeing the quality-of-service (QoS) of the downlink communication users. As a step further, several novel waveform designs have been proposed in \cite{8288677} with the purpose of minimizing the multi-user interference (MUI) for downlink communications under radar-specific constraints.
\\\indent It is noteworthy that in the DFRC scenarios, the targets to be detected might be potential eavesdroppers. This is most likely to appear in defense-related applications, where radar targets are usually adversary's combat platforms. In that case, it is highly possible that the critical information will be leaked to the radar targets by the emission of the DFRC waveform. Given the fact above, physical layer security (PHY-security) must be considered in DFRC designs \cite{8509094,8437135}. Particularly, it has been widely understood that the communication secrecy could be enhanced by exploiting the artificial noise (AN). For instance, in \cite{shu2018artificial}, the directional modulation method was adopted together with the AN to improve the communication performance within the direction of interest while degrading that of other directions. In \cite{li2013spatially}, the authors studied the AN-aided secrecy rate maximum problem with no structural restrictions on the AN in  multiple-input multiple-output (MISO) channel. Furthermore, in \cite{zhu2016linear}, AN based linear precoding designs have been proposed to ensure the secrecy performance in the system of massive  multiple-input multiple-output (MIMO). Aiming to guarantee the QoS of the downlink users while confusing the eavesdropper, AN has been exploited in \cite{liao2011qos} for transmit beamforming design. Nevertheless, to the best of our knowledge, little research efforts have been taken towards the direction of enhancing the PHY-security of the DFRC system.
\\\indent In this paper, we study the transmission security for the DFRC system, where the downlink cellular users are regarded as legitimate receivers, with the radar targets being regarded as potential eavesdroppers. Optimization problems are designed to guarantee the communication secrecy by minimizing the signal-to-interference-plus-noise ratio (SINR) at the target with the help of AN. In the meantime, the communication secrecy rate for the legitimate users is guaranteed by imposing SINR thresholds. By employing the assumption of  perfect communication CSI, the proposed optimization problems are designed under precise and uncertain knowledge about the location of the target, respectively, following with the complexity analysis of each algorithm. Finally, numerical results are provided by Monte-Carlo simulations, which show that the proposed method is capable of guaranteeing the secrecy performance while formulating a desired spatial beampattern towards the target.
\section{System Model}
We consider a dual-functional MIMO RadCom system as shown in Fig. 1. The BS is equipped with an $N$-antenna uniform linear array (ULA). It serves $K$ downlink single-antenna legitimate users while detecting targets at the same time. The targets are considered as potential eavesdroppers which may eavesdrop the information from the BS to legitimate users. For simplicity, we assume that there is a single target of interest in our system.
\begin{figure}
    \centering
    \includegraphics[width=0.95\linewidth]{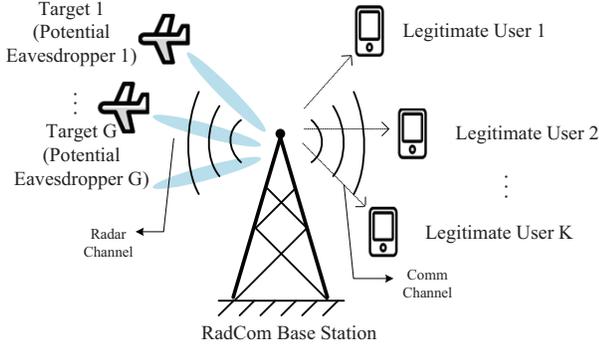}
    \caption{Dual-functional Radar-Communication system detecting targets which work as potential eavesdroppers.}
    \label{fig:1}
\end{figure}
\\\indent By exploiting the artificial noise in the beamforming design, the transmit matrix ${\mathbf{X}}$ can be expressed as
\begin{equation}\label{eq1}
    {\bf{X = WS + N}}\,
\end{equation}
where ${\mathbf{S}}=\left[ {{{\mathbf{s}}_1}, \cdots ,{{\mathbf{s}}_L}} \right]\in {\mathbb{C}^{K\times L}}$ is the desired signal from BS; $\mathbf{W} = {\left[ {{{\mathbf{w}}_1},{{\mathbf{w}}_2},...,{{\mathbf{w}}_K}} \right]^T}\in {\mathbb{C}^{N\times K}}$is the beamforming matrix; ${\mathbf{N}} \in {\mathbb{C}^{N\times L}}$ is an artificial noise matrix generated artificially for avoiding leaking information to targets. Without the loss of generality, we assume $\mathbb{E}\left[ {{\mathbf{s}}_{l}}\mathbf{s}_{l}^{H} \right]={\mathbf{I}}$, where ${{\mathbf{s}}_{l}}$ is the desired signal vector in the $l$-th time sloteq3. It is assumed that ${{\mathbf{n}}_i}\sim\mathcal{C}\mathcal{N}\left( {0,{{\mathbf{R}}_N}} \right),\forall i$, where ${{\mathbf{n}}_i}$ is \emph{i}-th vector of $\bf{N}$. ${{\mathbf{R}}_N}$ is the covariance matrix of the artificial noise. It follows that the covariance matrix of transmitted waveform can be written as
\begin{equation}\label{eq2}
    {{\mathbf{R}}_X} = \frac{1}{L}{\mathbf{X}}{{\mathbf{X}}^H} = \sum\limits_{i = 1}^K {{{\mathbf{W}}_i} + {{\mathbf{R}}_N}}.
\end{equation}
where ${{\mathbf{W}}_{i}}\triangleq{{\mathbf{w}}_{i}}\mathbf{w}_{i}^{H}$. $\mathbb{E}\left\{ \cdot  \right\}$ denotes the statistical expectation and ${{\left( \cdot  \right)}^{H}}$ represents the Hermitian transpose.
\\\indent The received symbol at legitimate users is given as
\begin{equation}\label{eq3}
    {\bf{Y = HX + Z}}\,
\end{equation}
where $\mathbf{H} \in {\mathbb{C}^{K \times N}}$ is the channel matrix; ${\mathbf{X}} \in {\mathbb{C}^{N \times L}}$ is the transmitted signal matrix, with $L$ being the length of the radar pulse/communication frame, ${\mathbf{Z}}\in {\mathbb{C}^{K \times L}}$ is the noise matrix, with ${{\mathbf{z}}_i}\sim\mathcal{C}\mathcal{N}\left( {0,{\sigma^2}{\mathbf{I}}_N} \right),\forall i$.
\\\indent To evaluate the performance of the system, several performance metrics are employed in this paper. Firstly, the SINR of the \emph{i}-th legitimate user is
\begin{equation}\label{eq4}
\begin{aligned}
    {{\text{SINR}}_{i}}&=\frac{\mathbb{E}\left[ {{\left| \mathbf{h}_{i}^{H}{{\mathbf{w}}_{i}}{{\mathbf{s}}_{i}} \right|}^{2}} \right]}{\sum\nolimits_{k\ne i, k=1}^{K}{\mathbb{E}\left[ {{\left| \mathbf{h}_{i}^{H}{{\mathbf{w}}_{k}}{{\mathbf{s}}_{k}} \right|}^{2}} \right]+{{\left| \mathbf{h}_{i}^{H}{{\mathbf{N}}_{i}} \right|}^{2}}+\sigma^{2}}}\\&=\frac{\text{tr}\left( \mathbf{h}_{i}^{T}{{\mathbf{W}}_{i}}\mathbf{h}_{i}^{*} \right)}{\sum\nolimits_{k\ne i, k=1}^{K}{\text{tr}\left( \mathbf{h}_{i}^{T}{{\mathbf{W}}_{k}}\mathbf{h}_{i}^{*} \right)}+\text{tr}\left( \mathbf{h}_{i}^{T}{{\mathbf{R}}_{N}}\mathbf{h}_{i}^{*} \right)+\sigma^{2}},
\end{aligned}
\end{equation}
Accordingly, the SINR at the target is given as
\begin{equation}\label{eq5}
    \text{SIN}{{\text{R}}_{E}}=\frac{{{\left| \alpha  \right|}^{2}}{{\mathbf{a}}^{H}}\left( \theta  \right)\sum\nolimits_{i=1}^{K}{{{\mathbf{W}}_{i}}}\mathbf{a}\left( \theta  \right)}{{{\left| \alpha  \right|}^{2}}{{\mathbf{a}}^{H}}\left( \theta  \right){{\mathbf{R}}_{N}}\mathbf{a}\left( \theta  \right)+\sigma^{2}},
\end{equation}
where $\theta $ represents the azimuth angle of the target, $\mathbf{a}\left( \theta  \right)={{\left[ \begin{matrix}
   1 & {{e}^{j2\pi \Delta \sin \left( \theta  \right)}} & \cdots  & {{e}^{j2\pi \left( N-1 \right)\Delta \sin \left( \theta  \right)}}  \\
\end{matrix} \right]}^{T}}\in {{\mathbb{C}}^{N\times 1}}$ denotes the steering vector of the transmit antenna array; $\Delta$ is the antenna spacing between adjacent antennas being normalized by the signal wavelength. Following \cite{parsaeefard2015improving}, the achievable secrecy rate can be defined as
\begin{equation}\label{eq6}
    \text{SR} = \frac{1}{2}{\left[ {\mathop {\min }\limits_i {{R_C}_i} - {R_E}} \right]^ + },
\end{equation}
where ${{R_C}_i} = {\log _2}\left( {1 + {\text{SIN}}{{\text{R}}_i}} \right)$, ${R_E} = {\log _2}\left( {1 + {\text{SIN}}{{\text{R}}_E}} \right)$, ${\left[  \cdot  \right]^{\text{ + }}}$ denotes $\max \left\{ { \cdot ,0} \right\}$.
\section{Problem Formulation}

\subsection{Minimizing ${\text{SIN}}{{\text{R}}_E}$ With Perfect CSI and Precise Target Location}\label{AA}
In this subsection, we aim to ensure secrecy rate by minimizing the SINR at target and guarantee the SINR at legitimate user maintaining above a certain threshold. Note that before designing the beamforming matrix and the artificial noise, an ideal radar beampattern should be obtained as the benchmark, which can be given by solving the following constrained least-squares (LS) problem \cite{fuhrmann2008transmit}
\begin{equation}\label{eq7}
\begin{gathered}
  \mathop {\min }\limits_{\eta ,{{\mathbf{R}}_d}}\; {\sum\limits_{m = 1}^M {\left| {\eta {P_d}\left( {{\theta _m}} \right) - {{\mathbf{a}}^H}\left( {{\theta _m}} \right){{\mathbf{R}}_d}{\mathbf{a}}\left( {{\theta _m}} \right)} \right|} ^2} \hfill \\
  s.t.\;\;\;{\text{tr}}\left( {{{\mathbf{R}}_d}} \right) = {P_0}, \hfill \\
  \;\;\;\;\;\;\;\;{{\mathbf{R}}_d} \succeq 0,{{\mathbf{R}}_d} = {\mathbf{R}}_d^H, \hfill \\
  \;\;\;\;\;\;\;\;\eta  \geqslant 0, \hfill \\
\end{gathered}
\end{equation}
where $\eta$ is a scaling factor; ${P_0}$ is the transmission power budget, $\left\{ {{\theta _m}} \right\}_{m = 1}^M$ denotes an angular grid covering the detection angle range in $\left[ {{ - }\pi /2,\pi /2} \right]$, ${{\mathbf{a}}\left( {{\theta _m}} \right)}$ denotes steering vector, ${{P_d}\left( {{\theta _m}} \right)}$ is the desired ideal beampattern gain at ${\theta _m}$, ${{{\mathbf{R}}_d}}$ represents the desired waveform covariance matrix. Given ${{\mathbf{R}}_d}$, our problem can be formulated as
\begin{subequations}\label{eq8}
\begin{align}
  &\mathop {\min }\limits_{{{\mathbf{W}}_i},{{\mathbf{R}}_N}} \frac{{{\left| \alpha  \right|}^{2}}{{\mathbf{a}}^{H}}\left( \theta_0  \right)\sum\nolimits_{i=1}^{K}{{{\mathbf{W}}_{i}}}\mathbf{a}\left( \theta_0  \right)}{{{\left| \alpha  \right|}^{2}}{{\mathbf{a}}^{H}}\left( \theta_0  \right){{\mathbf{R}}_{N}}\mathbf{a}\left( \theta_0  \right)+\sigma^{2}}, \forall i \hfill \\
  &s.t.\;\;\;{\left\| {{{\mathbf{R}}_X} - {{\mathbf{R}}_d}} \right\|^2} \leqslant {\gamma _{bp}}, \hfill \\
  &\;\;\;\;\;\;\;\;{\text{SIN}}{{\text{R}}_i} \geqslant {\gamma _b}, \forall i, \hfill \\
  &\;\;\;\;\;\;\;\;{P_t} = {P_0}, \hfill \\
  &\;\;\;\;\;\;\;\;{{\mathbf{W}}_i} = {\mathbf{W}}_i^H, {{\mathbf{W}}_i} \succeq 0, \forall i, \hfill \\
  &\;\;\;\;\;\;\;\;{\text{rank}}\left( {{{\mathbf{W}}_i}} \right) = 1, \forall i, \hfill \\
  &\;\;\;\;\;\;\;\;{{\mathbf{R}}_N} = {\mathbf{R}}_N^H,{{\mathbf{R}}_N} \succeq 0,
\end{align}
\end{subequations}
where $\theta_0$ denotes the location of targets known at the BS, ${\gamma _{bp}}$ is a pre-defined threshold that constrains the mismatch between the designed covariance matrix ${{{\mathbf{R}}_X}}$ and the desired ${{{\mathbf{R}}_d}}$. ${\gamma _{b}}$ denotes the predefined SINR threshold of each legitimate user, and finally ${\text{rank}}\left(  \cdot  \right)$ is rank operator.

\subsection{Minimizing ${\text{SIN}}{{\text{R}}_E}$ With Perfect CSI and Target Location Uncertainty}
In this subsection, we consider the case that the target's position is roughly known within the angular interval $\Phi=\left[\theta_0-\Delta\theta,\theta_0+\Delta\theta\right]$ due to the uncertainty in the target parameter estimation. To guarantee the secrecy rate from transmitter to legitimate users, the objective function is reformulated to minimize the sum of target's SINR at the possible locations in the angular interval aforementioned. To this end, wider beam needs to be formulated towards the uncertain interval to avoid missing the target. Inspired by the 3dB beampattern design approach in \cite{li2007mimo}, in our problem, we formulate the beampattern aiming to keep the power equivalent in the angular interval where the target is estimated to locate at. The proposed optimization problem can be formulated as
\begin{subequations}\label{eq9}
\begin{align}
  &\mathop {\min }\limits_{{{\mathbf{W}}_i},{{\mathbf{R}}_N}} \sum\limits_{{\theta _m} \in \Phi } {\frac{{{{\left| \alpha  \right|}^2}{{\mathbf{a}}^H}\left( {{\theta _m}} \right)\sum\nolimits_{i = 1}^K {{{\mathbf{W}}_i}} {\mathbf{a}}\left( {{\theta _m}} \right)}}{{{{\left| \alpha  \right|}^2}{{\mathbf{a}}^H}\left( {{\theta _m}} \right){{\mathbf{R}}_N}{\mathbf{a}}\left( {{\theta _m}} \right) + {\sigma ^2}}}} ,\forall i \hfill \\
  &s.t.\;{{\mathbf{a}}^H}\left( {{\theta _0}} \right){{\mathbf{R}}_X}{\mathbf{a}}\left( {{\theta _0}} \right) - {{\mathbf{a}}^H}\left( {{\theta _m}} \right){{\mathbf{R}}_X}{\mathbf{a}}\left( {{\theta _m}} \right) \geqslant {\gamma _s},\\
  &\;\;\;\;\;\;\forall {\theta _m} \in \Omega  \notag \hfill \\
  &\;\;\;\;\;\;{{\mathbf{a}}^H}\left( {{\theta _k}} \right){{\mathbf{R}}_X}{\mathbf{a}}\left( {{\theta _k}} \right) \leqslant \left( {1 + \alpha } \right){{\mathbf{a}}^H}\left( {{\theta _0}} \right){{\mathbf{R}}_X}{\mathbf{a}}\left( {{\theta _0}} \right), \\
  &\;\;\;\;\;\;\forall {\theta _k} \in \Phi \notag \hfill \\
  &\;\;\;\;\;\;\left( {1 - \alpha } \right){{\mathbf{a}}^H}\left( {{\theta _0}} \right){{\mathbf{R}}_X}{\mathbf{a}}\left( {{\theta _0}} \right) \leqslant {{\mathbf{a}}^H}\left( {{\theta _k}} \right){{\mathbf{R}}_X}{\mathbf{a}}\left( {{\theta _k}} \right),\\ &\;\;\;\;\;\;\forall {\theta _k} \in \Phi  \notag \hfill \\
  &\;\;\;\;\;\;{\text{SIN}}{{\text{R}}_i} \geqslant {\gamma _b}, \forall i, \hfill \\
  &\;\;\;\;\;\;{P_t} = {P_0}, \forall i, \hfill \\
  &\;\;\;\;\;\;{{\mathbf{W}}_i} = {\mathbf{W}}_i^H, \; {{\mathbf{W}}_i} \succeq 0, \forall i, \hfill \\
  &\;\;\;\;\;\;{\text{rank}}\left( {{{\mathbf{W}}_i}} \right) = 1,\forall i, \hfill \\
  &\;\;\;\;\;\;{{\mathbf{R}}_N} = {\mathbf{R}}_N^H,\; {{\mathbf{R}}_N} \succeq 0,
\end{align}
\end{subequations}
where ${{\theta _0}}$ is the main-beam location, ${\Omega}$ denotes the sidelobe region of interest, ${\gamma _s}$ is a bound of sidelobe power.
\section{The Proposed Approach}
\subsection{Approaches to Problem (8) and (9)}
In this section, we propose an iterative algorithm to solve the above optimization problems. Initially, it is straightforward to see that (7) is convex, which can be readily solved by use of standard numerical tools, such as CVX. According to \cite{shen2018fractional}, the problem (8) and (9) presented in section \uppercase\expandafter{\romannumeral3} can be both regarded as fraction programming (FP) problem. Let us denote
\begin{align*}
    &\operatorname{M}={\left| \alpha  \right|^2}{{\mathbf{a}}^H}\left( \theta_0  \right)\sum\nolimits_{i = 1}^K {{{\mathbf{W}}_i}} {\mathbf{a}}\left( \theta_0  \right),\forall i, \hfill \\
    &\operatorname{N}  = {\left| \alpha  \right|^2}{{\mathbf{a}}^H}\left( \theta_0  \right){{\mathbf{R}}_N}{\mathbf{a}}\left( \theta_0  \right) + {\sigma ^2}
\end{align*}
By the above notations, problem (8) can be relaxed in a convex form as
\begin{equation}\label{eq10}
\begin{aligned}
  &\mathop {\min }\limits_{{{\mathbf{W}}_i},{{\mathbf{R}}_N}} \operatorname{M}  - c\operatorname{N} , \hfill \\
  &s.t.\;\;\;{\left\| {{{\mathbf{R}}_X} - {{\mathbf{R}}_d}} \right\|^2} \leqslant {\gamma _{bp}}, \hfill \\
  &\;\;\;\;\;\;\;\;{\text{SIN}}{{\text{R}}_i} \geqslant {\gamma _b}, \forall i, \hfill \\
  &\;\;\;\;\;\;\;\;{P_t} = {P_0}, \hfill \\
  &\;\;\;\;\;\;\;\;{{\mathbf{W}}_i} = {\mathbf{W}}_i^H, {{\mathbf{W}}_i} \succeq 0, \forall i, \hfill \\
  &\;\;\;\;\;\;\;\;{\text{rank}}\left( {{{\mathbf{W}}_i}} \right) = 1,\forall i, \hfill \\
  &\;\;\;\;\;\;\;\;{{\mathbf{R}}_N} = {\mathbf{R}}_N^H,{{\mathbf{R}}_N} \succeq 0,
\end{aligned}
\end{equation}
To solve the original problem (8), the scaling factor c needs to be updated in each iteration, yielding
\begin{equation}\label{eq11}
    c\left[ {t + 1} \right] = \frac{{\operatorname{M}\left[ t \right]}}{{\operatorname{N}\left[ t \right]}},
\end{equation}
where $t$ is the index of iteration. To solve (10), SDR technique can be adopted by omitting the rank-1 constraint. For clarity, we summarize the above procedure in Algorithm 1.
\renewcommand{\algorithmicrequire}{\textbf{Input:}}
\renewcommand{\algorithmicensure}{\textbf{Output:}}
\begin{algorithm}
\caption{Alogrithm for solving FP problem (8)}
\label{alg:1}
\begin{algorithmic}
    \REQUIRE ${\mathbf{H}},{\mathbf{a}}\left( \theta_0 \right)$, ${\sigma ^2},\alpha ,{\gamma _b},{P_0},ite{r_{max}} \geqslant 2$, ${\gamma _{bp}}$.
    \ENSURE ${\mathbf{W}}_i^{\left( {iter} \right)},{\mathbf{R}}_N^{\left( {iter} \right)},i = 1, \cdots ,K$.
    \STATE 1. Compute ${{\mathbf{R}}_d}$. Reformulate problem (8) by (10). Set the iteration threshold $\varepsilon > 0$. Initialize ${c^{\left( 0 \right)}},{c^{\left( 1 \right)}},\left| {{c^{\left( 1 \right)}} - {c^{\left( 0 \right)}}} \right| > \varepsilon $.
    \WHILE {$iter \leqslant ite{r_{max}}$ and $\left| {{c^{iter + 1}} - {c^{iter}}} \right| \geqslant \varepsilon$ }
    \STATE 2. Solve the new convex optimization problem.
    \STATE 3. Update $c$ by (11).
    \STATE 4. Get updated ${{\mathbf{W}}_i}, \forall i,$ and ${{\mathbf{R}}_N}$ by solving (10) using SDR.
    \STATE 5. $ iter = iter + 1$.
    \ENDWHILE
\STATE 6. Obtain approximate solutions by eigenvalue decomposition or Gaussian randomization.
\end{algorithmic}
\end{algorithm}
\\\indent According to \cite{shen2018fractional}, it can be easily proven that $c$ is non-decreasing during the iterations. Consequently, the convergence of Algorithm 1 can be guaranteed.
\\\indent Following the similar procedure, we denote
\begin{align*}
    &A\left( {{\theta _m}} \right) = {\left| \alpha  \right|^2}{{\mathbf{a}}^H}\left( {{\theta _m}} \right){{\mathbf{R}}_N}{\mathbf{a}}\left( {{\theta _m}} \right) + {\sigma ^2} \hfill \\
    &B\left( {{\theta _m}} \right) = {\left| \alpha  \right|^2}{{\mathbf{a}}^H}\left( {{\theta _m}} \right)\sum\nolimits_{i = 1}^K {{{\mathbf{W}}_i}} {\mathbf{a}}\left( {{\theta _m}} \right)
\end{align*}
Problem (9) can be rewritten in the form
\begin{equation}\label{eq12}
\begin{aligned}
    &\mathop {\max }\limits_{{{\mathbf{W}}_i},{{\mathbf{R}}_N},{\mathbf{y}}} \sum\limits_{{\theta _m} \in \Phi } {\left( {2{y_m}\sqrt {A\left( {{\theta _m}} \right)}  - y_m^2B\left( {{\theta _m}} \right)} \right)}  \\
    &s.t.\;\;{{\mathbf{a}}^H}\left( {{\theta _0}} \right){{\mathbf{R}}_X}{\mathbf{a}}\left( {{\theta _0}} \right) - {{\mathbf{a}}^H}\left( {{\theta _m}} \right){{\mathbf{R}}_X}{\mathbf{a}}\left( {{\theta _m}} \right) \geqslant {\gamma _s},\forall {\theta _m} \in \Omega  \hfill \\
    &\;\;\;\;\;\;{{\mathbf{a}}^H}\left( {{\theta _k}} \right){{\mathbf{R}}_X}{\mathbf{a}}\left( {{\theta _k}} \right) \leqslant \left( {1 + \alpha } \right){{\mathbf{a}}^H}\left( {{\theta _0}} \right){{\mathbf{R}}_X}{\mathbf{a}}\left( {{\theta _0}} \right), \forall {\theta _k} \in \Phi \hfill \\
    &\;\;\;\;\;\;\left( {1 - \alpha } \right){{\mathbf{a}}^H}\left( {{\theta _0}} \right){{\mathbf{R}}_X}{\mathbf{a}}\left( {{\theta _0}} \right) \leqslant {{\mathbf{a}}^H}\left( {{\theta _k}} \right){{\mathbf{R}}_X}{\mathbf{a}}\left( {{\theta _k}} \right),\forall {\theta _k} \in \Phi  \hfill \\
    &\;\;\;\;\;\;\;{\text{SIN}}{{\text{R}}_i} \geqslant {\gamma _b}, \forall i, \hfill \\
    &\;\;\;\;\;\;\;{P_t} = {P_0}, \forall i, \hfill \\
    &\;\;\;\;\;\;\;{{\mathbf{W}}_i} = {\mathbf{W}}_i^H, \; {{\mathbf{W}}_i} \succeq 0, \forall i, \hfill \\
    &\;\;\;\;\;\;\;{\text{rank}}\left( {{{\mathbf{W}}_i}} \right) = 1,\forall i, \hfill \\
    &\;\;\;\;\;\;\;{{\mathbf{R}}_N} = {\mathbf{R}}_N^H,\; {{\mathbf{R}}_N} \succeq 0.
\end{aligned}
\end{equation}
Let ${\mathbf{y}}$ denote a collection of variables $\left\{ {{y_1}, \cdots ,{y_M}} \right\}$, where ${y_m}$ is updated iteratively by the following closed form when $\theta_m$ is fixed
\begin{equation}\label{eq13}
    y_m^* = \frac{{\sqrt {A\left( {{\theta _m}} \right)} }}{{B\left( {{\theta _m}} \right)}}.
\end{equation}
The problem (12) can be solved again by the SDR technique. We note that eigenvalue decomposition or Gaussian randomization is required to obtain an approximate solution. For clarity, the above procedure is summarized in Algorithm 2.
\renewcommand{\algorithmicrequire}{\textbf{Input:}}
\renewcommand{\algorithmicensure}{\textbf{Output:}}
\begin{algorithm}
\caption{Alogrithm for solving FP problem (9)}
\label{alg:2}
\begin{algorithmic}
    \REQUIRE ${\mathbf{H}},{\mathbf{a}}\left( \theta  \right)$ or ${\mathbf{a}}\left( \theta_m  \right)$, ${\sigma ^2},\alpha ,{\gamma _b},{P_0},ite{r_{max}} \geqslant 2$, $\Delta\theta$.
    \ENSURE ${\mathbf{W}}_i^{\left( {iter} \right)},{\mathbf{R}}_N^{\left( {iter} \right)},i = 1, \cdots ,K$.
    \STATE 1. Compute ${{\mathbf{R}}_d}$. Reformulate problem (9) by (12). Set the iteration threshold $\varepsilon > 0$.
    \WHILE {$iter \leqslant ite{r_{max}}$ and $\left\| {{{\mathbf{y}}^{iter + 1}} - {{\mathbf{y}}^{iter}}} \right\| \geqslant \varepsilon $}
    \STATE 2. Solve the new convex optimization problem.
    \STATE 3. Update ${\mathbf{y}}$ by (13).
    \STATE 4. Get updated ${{\mathbf{W}}_i}, \forall i,$ and ${{\mathbf{R}}_N}$ by solving (12) using SDR.
    \STATE 5. $ iter = iter + 1$.
    \ENDWHILE
\STATE 6. Obtain approximate solutions by eigenvalue decomposition or Gaussian randomization.
\end{algorithmic}
\end{algorithm}
\subsection{Complexity Analysis}
In this subsection, we analyze the computational complexity of problem (8) and (9), both of which have been transformed to SDP problem solved by interior point method (IPM) \cite{6891348}. Given $\epsilon>0$, we obtain an $\epsilon$-optimal solution after the required number of iterations. In the first optimization problem, to generate the ideal beampattern ${{\mathbf{R}}_d}$ in (7), which is a typical SDP problem including a linear matrix inequality (LMI) constraint of size $2N$, two LMI constraints of size $N$, and an LMI constraint of size 1, so the complexity is $\mathcal{O}\left( {\sqrt {4N + 1} \left( {4{N^6} + 3{N^5} + {N^4} + {N^2}} \right)} \ln \left( {{1 \mathord{\left/
 {\vphantom {1 \varepsilon }} \right.\kern-\nulldelimiterspace} \epsilon }} \right)\right)$. It is notable that problem (8) involves both LMI and second-order cone (SOC) constraints, which contains $2K+1$ LMI constraints of size $N$, $K+1$ LMI constraints of size $2N$, an LMI constraint of size 1 and an SOC constrain of size $N$. Accordingly, the computational complexity in (8) can be given in (14).
\begin{figure*}
\begin{equation}
\begin{aligned}
    &\mathcal{O}\left( {{N_{iter}}\ln \left( {{1 \mathord{\left/
    {\vphantom {1 \varepsilon }} \right.
    \kern-\nulldelimiterspace} \varepsilon }} \right)\sqrt {2N\left( {K + 1} \right) + K + 3}  \cdot K{N^2}\left( {\left( {K + 1} \right)\left( {K{N^2} + 1} \right) + 2{N^3}\left( {{K^2}N + KN + K + 1} \right)} \right)} \right)\\
    &+\mathcal{O}\left( {{N_{iter}}\ln \left( {{1 \mathord{\left/
    {\vphantom {1 \varepsilon }} \right.
    \kern-\nulldelimiterspace} \varepsilon }} \right)\sqrt {2N\left( {K + 1} \right) + K + 3}  \cdot K{N^4}\left( {{K^2}{N^2} + 1} \right)} \right)+\mathcal{O}\left( {\left( {K + 1} \right){N^3}} \right)
\end{aligned}
\end{equation}
\begin{equation}
\begin{aligned}
    &\mathcal{O}\left( {{N_{iter}}\ln \left( {{1 \mathord{\left/
     {\vphantom {1 \varepsilon }} \right.
    \kern-\nulldelimiterspace} \varepsilon }} \right)\sqrt {2N\left( {K + 1} \right) + K + {\Omega _0} + 2{\Phi _0} + 1}  \cdot K{N^2}\left( {K{N^2} + 1} \right)\left( {K + {\Omega _0} + 2{\Phi _0} + 1} \right)} \right)\\
    &+\mathcal{O}\left( {{N_{iter}}\ln \left( {{1 \mathord{\left/
    {\vphantom {1 \varepsilon }} \right.
    \kern-\nulldelimiterspace} \varepsilon }} \right)\sqrt {2N\left( {K + 1} \right) + K + {\Omega _0} + 2{\Phi _0} + 1}  \cdot K{N^2}\left( {2{N^3}\left( {{K^2}N + KN + K + 1} \right) + {K^2}{N^4}} \right)} \right)\\
    &+\mathcal{O}\left( {\left( {K + 1} \right){N^3}} \right)
\end{aligned}
\end{equation}
\hrulefill
\end{figure*}
 We note that $\mathcal{O}\left( {\left( {K + 1} \right){N^3}} \right)$ is the complexity of eigenvalue decomposition\footnote{We adopt eigenvalue decomposition method to get the approximate result because of the high complexity of Gaussian randomization.}. For simplicity, the computational complexity can be commonly given as ${\mathcal{O}}\left( {N_{iter}}{{K^{3.5}}{N^{6.5}}\ln \left( {{1 \mathord{\left/
 {\vphantom {1 \varepsilon }} \right.
 \kern-\nulldelimiterspace} \epsilon }} \right)} \right)+\mathcal{O}\left( {\left( {K + 1} \right){N^3}} \right)$, where $N_{iter}$ represents iteration times.
\\\indent Similarly, in problem (9), we regard both $\Omega$ and $\Phi$ as discrete domains which represent a collection of angles region. This problem involves LMI constraints only. Specifically, it includes $K+1$ LMI constraints of size $2N$, $2K+2\Phi+\Omega+1$ LMI constraints of size $N$, an LMI constraints of size 1. The iteration complexity is shown in (15).
 Then, the computational complexity can be approximately given as ${\mathcal{O}}\left( {N_{iter}} {{K^{3.5}}{N^{6.5}}\ln \left( {{1 \mathord{\left/
 {\vphantom {1 \varepsilon }} \right.
 \kern-\nulldelimiterspace} \epsilon }} \right)} \right)+\mathcal{O}\left( {\left( {K + 1} \right){N^3}} \right)$.
\section{Simulation Results}
\begin{figure}
    \centering
    \includegraphics[width=0.95\linewidth]{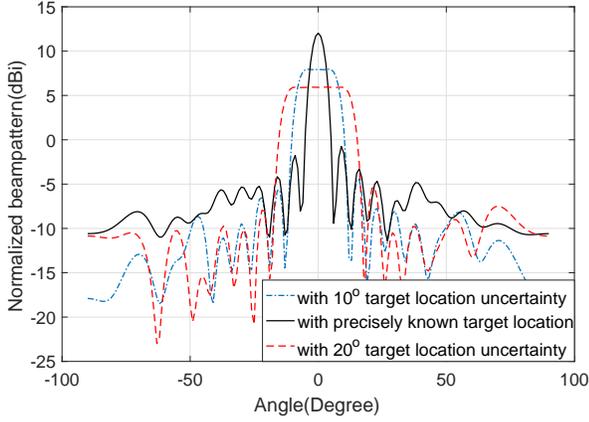}
    \caption{Radar beampattern obtained on different circumstances, $N = 18,K = 4,{\gamma _b} = 10, {P_0} = 30{\text{dBm}}$.}
    \label{Fig.2}
\end{figure}
\begin{figure}
    \centering
    \includegraphics[width=0.95\linewidth]{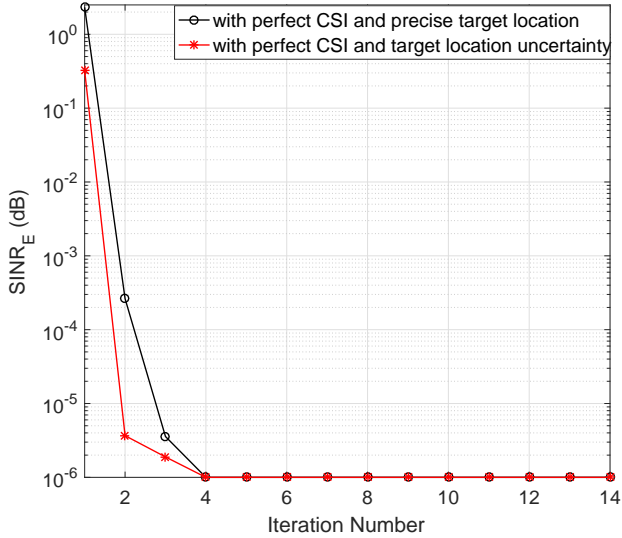}
    \caption{SINR at eavesdropper with iteration number variance, $N = 18,K = 4,{\gamma _b} = 10$, ${P_0} = 30{\text{dBm}}$.}
    \label{Fig.3}
\end{figure}
\begin{figure}
    \centering
    \includegraphics[width=0.95\linewidth]{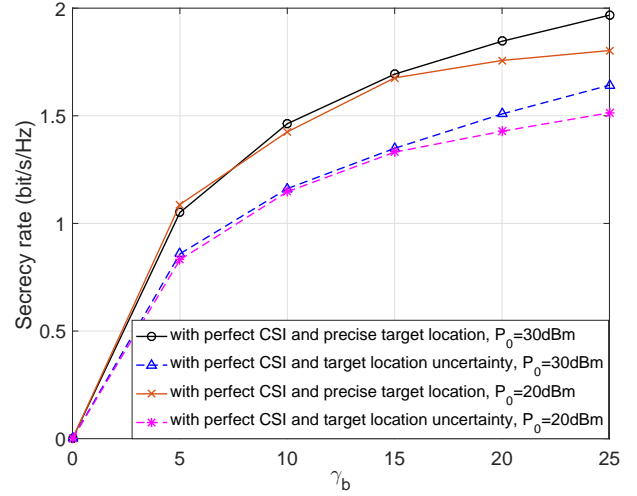}
    \caption{Achievable secrecy rate versus the threshold of SINR at legitimate users, $N = 18,K = 4$.}
    \label{Fig.4}
\end{figure}
\begin{figure}
    \centering
    \includegraphics[width=0.95\linewidth]{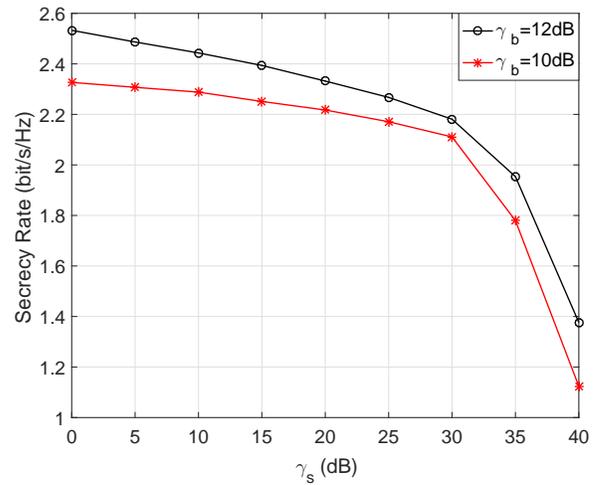}
    \caption{Achievable secrecy rate versus the threshold of sidelobe power with different SINR threshold at legitimate users, $N = 18,K = 4, {P_0} = 30{\text{dBm}}$.}
    \label{Fig.5}
\end{figure}
In this section, we present the simulation results of proposed methods. For all simulations, we employ a ULA with half-wavelength spacing between adjacent antennas. The length of communication frame/radar pulse is set as $L=30$. Without loss of generality, each entry of the channel matrix ${\mathbf{H}}$ is assumed to obey standard Complex Gaussian distribution, i.e., ${h_{i,j}} \sim {\mathcal{CN}}\left( {0,1} \right)$.
\\\indent We first show in Fig. 2 the obtained radar beampatterns with precise and uncertain target angles, which are formulated by solving problem (8) and (9) respectively. We set $\Delta\theta=5^\circ$ and $\Delta\theta=10^\circ$ when the target location is known roughly, which represented by dashed.For the case of precise target angle, it is noted that a narrow beam is formulated towards the direction of interest. For the cases with uncertain target angles, on the other hand, the mainbeam power decreases with the angle of location uncertainty being broadened, which demonstrates the tradeoff between the power of beampattern pointing to the target location and the precision of target location known at transmitter.
\\\indent The convergence of SINR at eavesdropper is demonstrated in Fig. 3. It is obvious that both Algorithm 1 and 2 converge to the optimum within a modest number of iterations.
\\\indent The secrecy rate of the legitimate communication link in terms of a rising SINR threshold $\gamma_b$ is shown in Fig. 4, with $\gamma_{bp} = 60$, $\gamma_s = 10^4$. It can be observed that the secrecy rate increases with the growth of the SINR threshold at legitimate users. It is noteworthy that the secrecy rate for the case with precisely known target location  is higher than that of the case with  uncertain target location. Additionally, the secrecy rate reduces with the descending tendency of the power budget.
\\\indent Finally, Fig. 5 represents the secrecy rate performance versus the given threshold of sidelobe $\gamma_s$, with ${P_0} = 30{\text{dBm}}, \Delta\theta=5^\circ$, and the SINR threshold at users is given as ${\gamma _b} = 10{\text{dB}}$ and ${\gamma _b} = 20{\text{dB}}$ respectively. It is notable that the secrecy rate decreases with the increasing of $\gamma_s$. It is notable that the decreasing tendency of secrecy rate gets obvious when $\gamma_s$ is greater than 30dB.
\section{Conclusion}
We have studied the AN-aided method to ensure the physical layer security in DFRC system. It is assumed that the communication CSI is perfectly known while the target location might be inaccurately estimated. To guarantee communication secrecy while detecting targets (which are potential eavesdroppers), we have minimized the SINR at the targets by formulating FP optimization problems, which can be equivalently recast as a series of sub-problems with convex objective functions. By dropping the rank-1 constraint, each sub-problem can be relaxed as an SDP, and can be thus solved by numerical tools. Numerical results have been provided to verify the effectiveness of the proposed approaches, which show that it it is feasible to guarantee both the performance of the radar beampattern and the communication secrecy by the proposed optimization based designs.
\section*{Acknowledgement}
This work has received funding from the the European Union's Horizon 2020 research and innovation programme under the Marie Sk\l odowska-Curie Grant Agreement No. 793345 and China Scholarship Council (CSC).

\bibliographystyle{IEEEtran}
\bibliography{IEEEabrv,CEP_REF}
\vspace{12pt}
\end{document}